\documentclass[aps,prl,groupedaddress,superscriptaddress,showpacs,reprint]{revtex4-1}
\usepackage{amsmath}
\usepackage{hyperref}
\usepackage{graphicx}
\usepackage{stackengine}
\usepackage{color}
\usepackage{subfigure}

\usepackage{helvet}
\usepackage{times}
\usepackage{enumitem}
\usepackage{ulem}

\begin{document}

\title{Unconditional and robust quantum metrological advantage beyond NOON states}

\author{Jian Qin}
\thanks{These authors contributed equally to this work.} 
\author{Yu-Hao Deng}
\thanks{These authors contributed equally to this work.}
\author{Han-Sen Zhong}
\thanks{These authors contributed equally to this work.}
\author{Li-Chao Peng}
\author{Hao Su}
\author{Yi-Han Luo}
\author{Jia-Min Xu}
\author{Dian Wu}
\author{Si-Qiu Gong}
\author{Hua-Liang Liu}
\author{Hui Wang}
\author{Ming-Cheng Chen}
\author{Li Li}
\author{Nai-Le Liu}

\author{Chao-Yang Lu}
\author{Jian-Wei Pan}

\affiliation{Hefei National Research Center for Physical Sciences at the Microscale and School of Physical Sciences, University of Science and Technology of China, Hefei, Anhui 230026, China}
\affiliation{CAS Center for Excellence and Synergetic Innovation Center in Quantum Information and Quantum Physics, University of Science and Technology of China, Shanghai 201315, China}
\affiliation{Hefei National Laboratory, University of Science and Technology of China, Hefei 230088, China}

\date{\today}

\begin{abstract}
Quantum metrology employs quantum resources to enhance the measurement sensitivity beyond that can be achieved classically. While multi-photon entangled NOON states can in principle beat the shot-noise limit and reach the Heisenberg limit, high NOON states are difficult to prepare and fragile to photon loss which hinders it from reaching unconditional quantum metrological advantages. Here, we combine the idea of unconventional nonlinear interferometers and stimulated emission of squeezed light, previously developed for photonic quantum computer Jiuzhang, to propose and realize a new scheme that achieves a scalable, unconditional, and robust quantum metrological advantage. We observe a 5.8(1)-fold enhancement above the shot-noise limit in the Fisher information extracted per photon, without discounting for photon loss and imperfections, which outperforms ideal 5-NOON states. The Heisenberg-limited scaling, the robustness to external photon loss, and the ease-of-use of our method make it applicable in practical quantum metrology at low photon flux regime. 
\end{abstract}

\maketitle
 
In optical phase measurements, especially at regimes with low photon flux, it is of fundamental interest to maximize the Fisher information \cite{helstrom1969quantum} that can be extracted per photon. Given finite resources, that is, the total number of photons that traverse the sample, it has been shown that the phase sensitivity is bound to shot-noise limit (SNL) using classical light \cite{giovannetti2004quantumenhanced,polino2020photonic,braunstein1992quantum}. Quantum resources can be employed to achieve phase sensitivity beyond the SNL \cite{giovannetti2004quantumenhanced}, which is called \textit{supersensitivity}. In 1981, Caves proposed the first phase supersensitive measurement protocol using squeezed light \cite{caves1981quantummechanical}, which were demonstrated experimentally \cite{xiao1987precision,grangier1987squeezedlight} later and used in GEO600 \cite{the2011gravitational} and the Advanced Laser Interferometer Gravitational-Wave Observatory \cite{tse2019quantum,acernese2019increasing} recently.
  
In quantum mechanics, the Heisenberg uncertainty principle places a fundamental limit for sensitivity \cite{giovannetti2004quantumenhanced,polino2020photonic}. It has been shown that a multi-photon path-entangled state, so called NOON state \cite{lee2002quantum}, can achieve the Heisenberg limit in principle. The interference fringes of the N-photon NOON states have a period N times shorter than using single photons \cite{mitchell2004super,walther2004broglie}, a phenomena called super-resolution. The phase super-resolution has been demonstrated using probabilistic and post-selected NOON states with up to 5 photons \cite{mitchell2004super,nagata2007beating,afek2010highnoon,matthews2009manipulation,rozema2014scalable} and multi-photon entangled states \cite{polino2020photonic,walther2004broglie,resch2007timereversal,okamoto2008beating,xiang2011entanglement,wang201818,zhong201812photon}.

However, it should be noted that the super-resolution is not equivalent to the super-sensitivity \cite{resch2007timereversal}. An unconditional quantum metrological advantage is achieved if the measured sensitivity per resource beats the SNL when all the used quantum resources are taken into account. For example, the total number of photons effectively employed in the experiment should not be corrected by imperfections such as photon loss, state fidelity, detection, and post-selection. The only unconditional violation of SNL at the low photon flux regime \cite{slussarenko2017unconditional} was demonstrated in 2017 but that scheme was limited for a 2-photon NOON state.

Unconditional quantum metrological advantages beyond the 2-photon NOON-states remained challenging. One reason is that the multi-photon NOON-state based super-sensitivity is very fragile to the photon loss. Even a small amount of photon loss will balance out the quantum-gained sensitivity \cite{demkowicz2009quantum}. However, high-efficiency preparations of high-NOON states without post-selection, as well as high-efficiency output photon projection and detection, have been long-standing challenges in the field over decades \cite{resch2007timereversal,datta2011quantum}. 

In a different path, Yurke \textit{et al.} proposed an unconventional interferometer to achieve the Heisenberg scaling employing cascaded optical parametric amplifiers, instead of the passive beam splitters \cite{yurke1986su}. Such nonlinear interferometers, also referred to as SU(1,1) interferometers, have been implemented using atomic four-wave mixing \cite{jing2011realization,hudelist2014quantum,gupta2018optimized}, and bulk nonlinear medium \cite{manceau2017detection,frascella2019wide}. However, these demonstrations haven't reached the unconditional advantage.

\begin{figure*}[!htp]
	\includegraphics[width=0.7\textwidth]{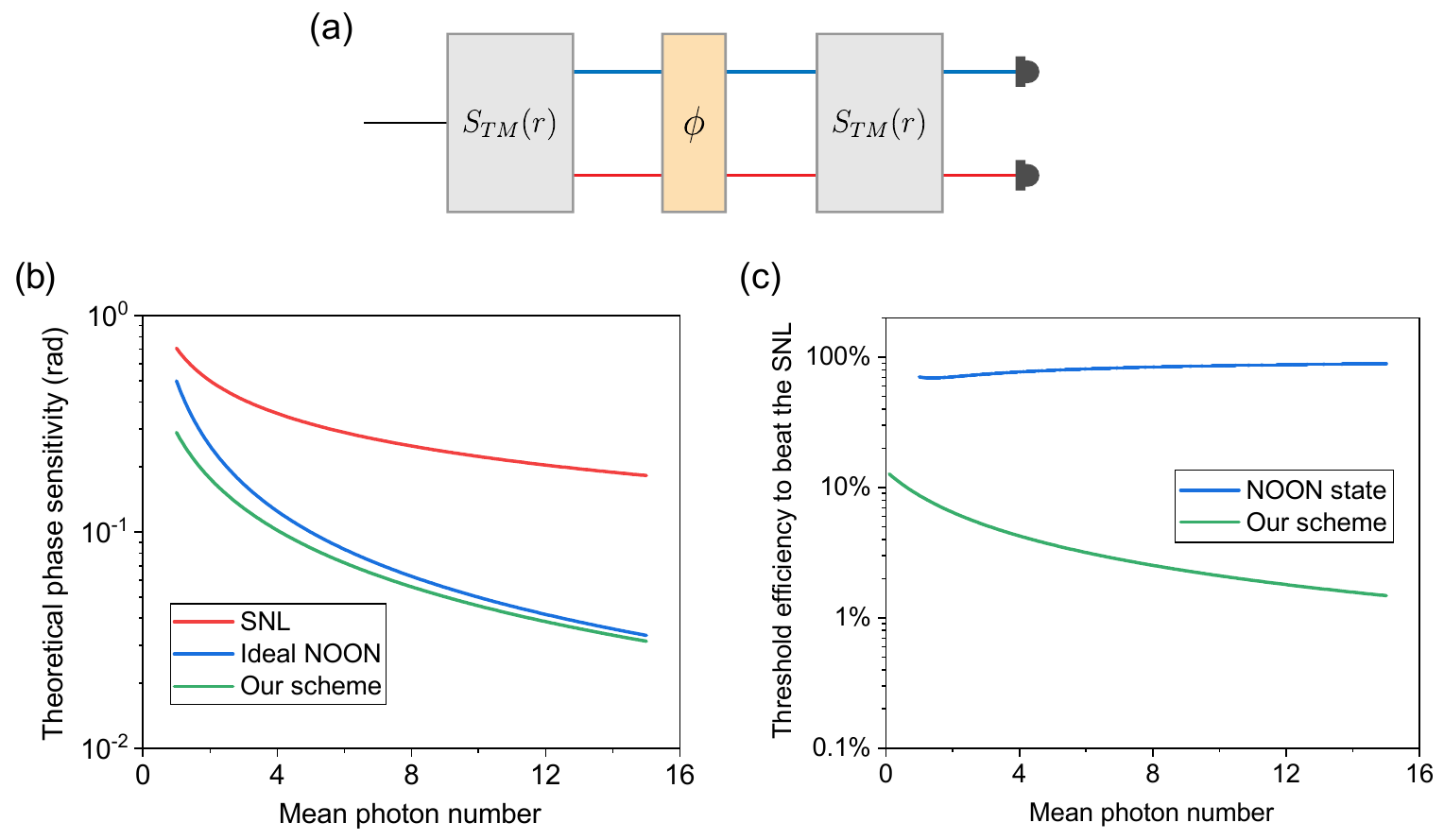}
	\caption{Principle of the stimulated squeezing nonlinear interferometer. (a) An input vacuum state is squeezed by a pair of two-mode squeezers ($S_{TM}$) with a relative phase $\phi$ which is to be measured. The output two-mode squeezed state is detected at single-photon level. (b) A comparison of theoretically predicted phase sensitivity for various measurement schemes assuming perfect visibility and efficiency. (c) A comparison of efficiency threshold to surpass the shot-noise limit (SNL) between our scheme and NOON-state at different mean photon number.}
	\label{fig:fig1}
\end{figure*}

In this Letter, for the first time, we combine the idea of nonlinear interferometer with stimulated emission of squeezed light \cite{zhong2021phaseprogrammable,lamas2001stimulated}, to demonstrate a scalable, unconditional, and robust quantum metrological advantage. A record-high Fisher information per photon, 11.6(1) $\rm{rad}^{-2}$, is directly observed, without discounting for any experimental imperfection. This not only unconditionally beats the SNL, but also surpass the limit that can be achieved using even ideal 5-NOON states. Using our method, we further demonstrate a case study for practical, real-time quantum-enhanced phase measurement at the low photon flux regime.

Figure \ref{fig:fig1} illustrates the working principle of the stimulated squeezing nonlinear interferometer which consists of a pair of two-mode squeezer ($S_{TM}$). The first $S_{TM}$ transforms an input vacuum state into a two-mode squeezed state (TMSS), which then acquires a to-be-measured phase $\phi$. The TMSS is sent to the second $S_{TM}$ and detected by two threshold single-photon detectors. The output state can be written as:
\begin{equation}
|\psi (\phi )\rangle =S_{TM}(r)  U(\phi )  S_{TM}(r)|0\rangle ,
\label{eq:eq1}
\end{equation}
where $S_{TM}(r)=e^{r\left( \hat{a}\hat{b}-\hat{a}^{\dagger}\hat{b}^{\dagger} \right) }$ is the $S_{TM}$ operator, $r$ is the squeezing parameter, and $U(\phi )=e^{i\left( \hat{a}^{\dagger}\hat{a}+\hat{b}^{\hat{\dagger}}\hat{b} \right) \phi}$  is the phase shift operator.

In the output, there are four possible outcomes: $p_{00}$, which refers to no click on both detectors, $p_{01}$($p_{10}$), which refers to click on the upper (lower) detector and no click on the other detector, and $p_{11}$ , which refers to coincident click on both detectors. We can calculate these probabilities and analyze the phase sensitivity as Fisher information per trial:
\begin{equation}
	F=\sum_{i,j=0,1}{\left( \frac{\partial \ln p_{ij}}{\partial \phi} \right) ^2}p_{ij}.
	\label{eq:fisher}
\end{equation}
The Fisher information has a maximum of $F_{\max}=4\sinh ^2\left( 2r \right)$. The mean photon number passing through by the phase sensor is $\bar{n}=2\sinh ^2r$. Therefore, the phase sensitivity could be expressed as:
\begin{equation}
	\Delta \phi =\frac{1}{\sqrt{F_{\max}}}=\frac{1}{2\sqrt{\bar{n}\left( \bar{n}+2 \right)}}.
\end{equation}
This equation confirms that the phase sensitivity of our scheme saturate the Heisenberg scaling.

Figure \ref{fig:fig1}b shows a comparison of the phase measurement sensitivity using Fig. \ref{fig:fig1}a with the SNL and the protocol using NOON states ideally (assuming a unity state fidelity and unity efficiencies in the generation, propagation and detection). The sensitivity scaling of our scheme is similar to the NOON state, which agrees with Eq. \ref{eq:fisher}, and clearly beats the SNL. Note that as a conservative comparison, we only count for the photons actually passing though the sample, which, for the case of an N-photon NOON states in a Mach-Zehnder interferometer configuration, the average photon number is $N/2$.

As Fig. \ref{fig:fig1}b consider the theoretical sensitivity only under perfect conditions, it is necessary to further analyze the performance under the most relevant realistic noise: the photon loss. For the N-photon NOON states to surpass the SNL, the threshold of the single-photon system efficiency is $\eta _{N00N}=\sqrt[N]{1/N}$, which asymptotically tends to 100\% and thus sets an unrealistically demanding experimental challenge (see the blue line in Fig. \ref{fig:fig1}c). The threshold detection efficiency of our scheme to surpass the SNL is $\eta _{TM}=1-\sqrt{1-1/2\left( \bar{n}+2 \right)}$ (see Supplemental Materials \cite{suppl} for a detailed derivation). As shown in the green line of Fig. \ref{fig:fig1}c, this is significantly lower than that of the NOON states \cite{manceau2017detection,plick2010coherentlightboosted,marino2012effect}. Moreover, in contrast to the NOON states, our threshold curve decrease as a function of the mean photon number, making it experimentally feasible and robust to the most sensitive noise in optical systems, the external photon loss (photon loss after the interferometer).

\begin{figure}[!htp]
	\includegraphics[width=0.48\textwidth]{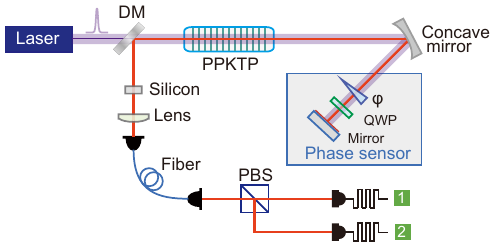}
	\caption{Experimental setup. A femtosecond pulsed laser is used to pump the periodically poled potassium titanyl phosphate (PPKTP) crystal as the squeezer. The concave mirror and reflection mirror form a compact 4-f optics to mode-match between both the pumping laser and squeezed photons of the first squeezing process to the second squeezing process. The relative phase between the two squeezing operation is tuned by a wedge plate. The quarter-wave plate (QWP) is used to exchange the polarization of the two-mode squeezed light for compensation of the birefringence walk-off. The two-mode squeezed photons are collected into single-mode fiber and detected by two superconducting nanowire single photon detectors. A dichromatic mirror (DM) and silicon plate are used to filter out the pumping laser from the squeezed photons.}  
	\label{fig:fig2}
\end{figure}

In our experiment, we use stimulated emission of TMSS, previously developed for photonic quantum computer, \textit{Jiuzhang} \cite{zhong2021phaseprogrammable}, to implement the nonlinear interferometer. The setup is shown in Fig. \ref{fig:fig2}. First, transform-limited laser pulses at a central wavelength of 775 nm are focused on a 4-mm-thickness periodically poled potassium titanyl phosphate (PPKTP) crystal to generate the TMSS at 1550 nm. The PPKTP is carefully designed to fulfill a collinear frequency-uncorrelated and degenerate type-\uppercase\expandafter{\romannumeral2} phase matching \cite{zhong2020quantum}. After the first pass, the pump laser and the collinear TMSS photons are reflected back and re-focused by a concave mirror, which are then used as seeds to stimulate the second parametric down-conversion process. Note that our experimental set-up naturally integrates the double-pass metrology protocol \cite{higgins2007entanglementfree}.

After filtering out the pump laser using a dichroic mirror and a silicon plate, the stimulated TMSS is collected into a single-mode fiber, projecting the two modes into the same spatial mode. The output is split by a polarizing beam splitter and detected by two superconducting nanowire single-photon detectors with a detection efficiency of $\sim$93\%. The PPKTP and the final planar mirror are placed at the focal points of the middle concave mirror to form a 4f optical system, which ensures the two parametric down-conversion processes have an optimal spatial matching. The birefringent walk-off between the horizontally and vertically polarization is compensated using a quarter-wave plate. The to-be-measured relative phase $\phi$ is added using an anti-reflection-coated wedge plate, where, due to material dispersion, the pumping laser and TMSS accumulate different phases. The $\phi$ can be tuned by a motorized translation stage.

\begin{figure*}[!htp]
	\includegraphics[width=0.97\textwidth]{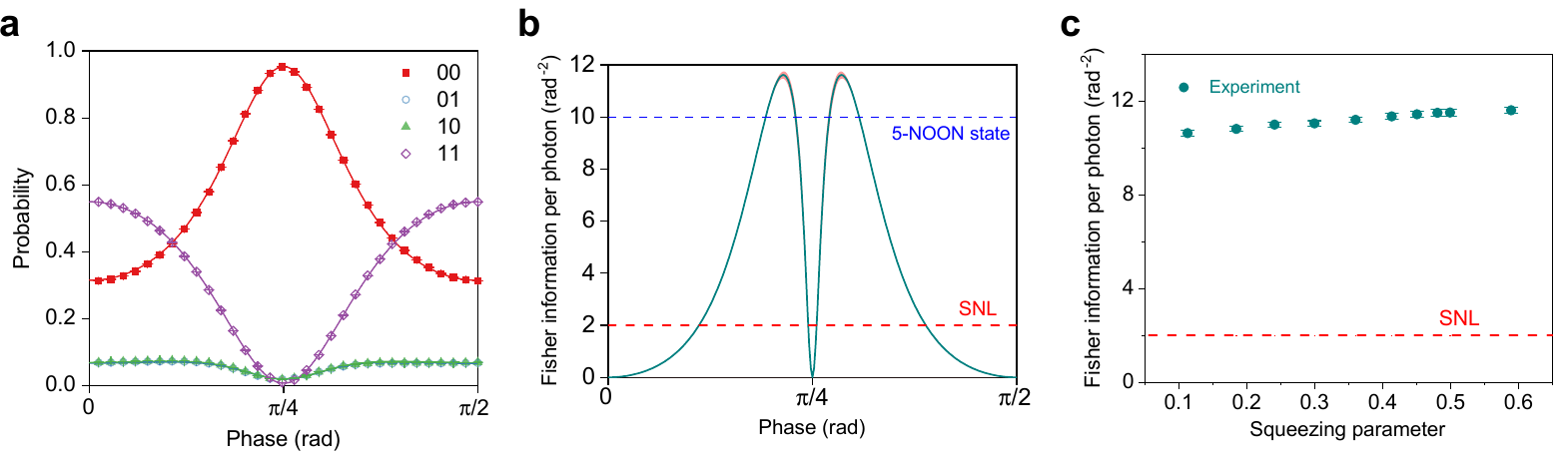}
	\caption{Experimentally measured output detection probability and the corresponding Fisher information. (a) The green (blue) points correspond to experimentally measured \{1(0), 0(1))\} coincidence events, the purple (red) points correspond to \{1(0), 1(0)\} coincidence events. Solid curves are fitted based on theoretical model (see Supplementary Materials \cite{suppl}). The error bar is smaller than the plot dot size. (b) The Fisher information per photon as a function of the phase. The cyan curve corresponds to the experimental results, while the red lines correspond to SNL. The shaded areas indicate 95\% confidence region. (c) The measured Fisher information per photon for various squeezing parameter and its comparison with SNL.}
	\label{fig:fig3}
\end{figure*}

In the first experiment, we choose a squeezing parameter of the TMSS to be 0.59(1) by tuning the power of the pumping laser, corresponding to a mean photon number of 0.78(1). The total system efficiency of the horizontal and vertical modes is 0.744(4) and 0.751(4), respectively. The squeezing parameters and the system efficiencies are carefully calibrated from the directly measured data, which is crucial for unconditional metrology advantage. The detailed calibration method is presented in Supplementary Materials \cite{suppl}. To measure the phase sensitivity, we tune the wedge plate to scan the phase from  0 to $\pi$, and record the four possible output signals  $p_{00}$, $p_{01}$, $p_{10}$ and $p_{11}$. The interference fringes of $p_{11}$ has a high visibility of 96.6(2)\% (Fig. \ref{fig:fig3}a), which reflect the degree of mode matching between the seed and the stimulated TMSS.

\begin{figure*}[!htp]
	\includegraphics[width=0.87\textwidth]{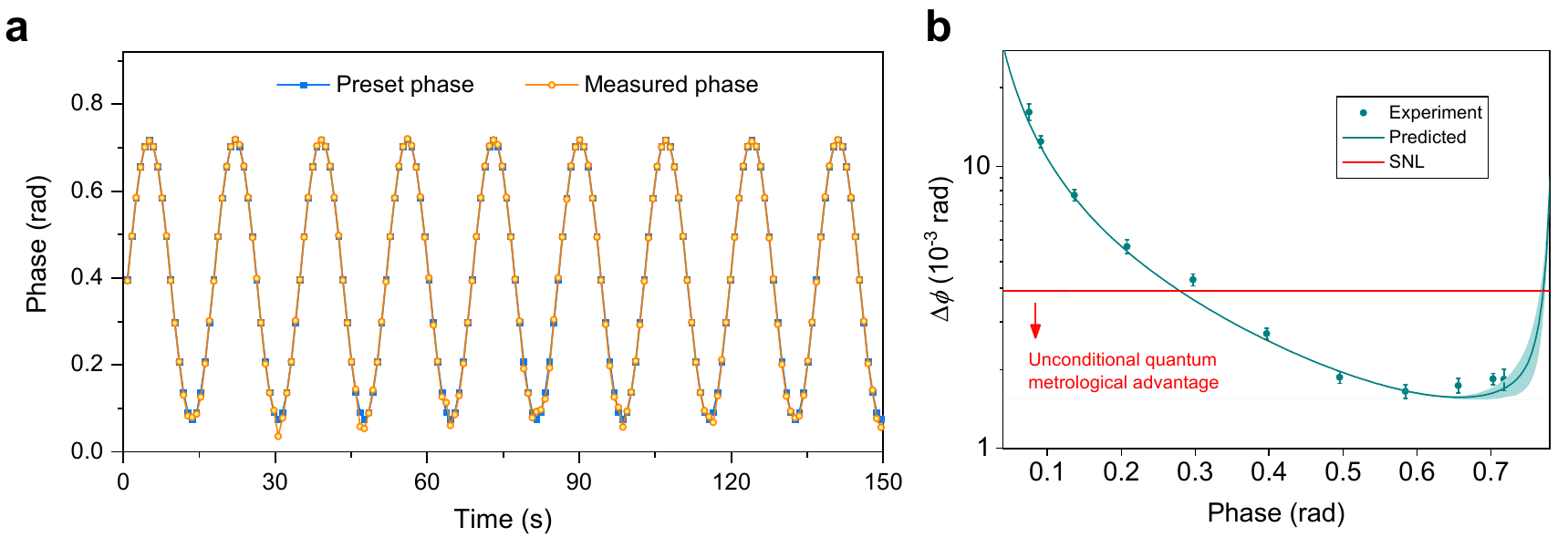}
	\caption{Phase sensitivity in a real-time measurement. (a) Dynamical tracing of the phase. The phase $\phi$ is tuned in steps with a motor-driven wedge plate for 150 seconds, and measured in real time with a sampling time of 0.2 second for each setting. The orange (blue) dots are measured (set) phases. (b) Experimentally measured phase uncertainty. Dots are the standard deviation of the measured phase corresponding to the data in (a), whose uncertainty is obtained by bootstrapping technique. The solid curve is the minimal standard deviation estimated from Fisher information. Red line corresponds to the SNL. Shaded areas correspond to 95\% confidence region.}
	\label{fig:fig4}
\end{figure*}

We use Fisher information per photon that traverse the sample to quantify the quantum metrological advantage at low photon flux regime. Following Eq. \ref{eq:fisher}, based on the interference fringes in Fig. \ref{fig:fig3}a, we extract the Fisher information per photon, which is shown in Fig. \ref{fig:fig3}b as a function of phase. The optimal Fisher information per photon reaches 11.6(1) $\rm{rad}^{-2}$ at the most sensitive phase points of 0.67 and 0.90. This is well above the SNL (red dash line), as shown in Fig. \ref{fig:fig3}b. It also exceeds the predicted value using an ideal 5-photon NOON state and perfect detectors, and arbitrary number of NOON states if using detectors with the same efficiency as in our work. Under different squeezing parameters at a range of 0.11(1) to 0.59(1), the Fisher information per photon is measured and plotted in Fig. \ref{fig:fig3}c where all data points exceed the SNL and 5-photon NOON states. In addition, the data shows a slight increase as a function of mean photon number, pointing a way to further improve the measurement sensitivity. With realistic improvements of the mode overlapping to 0.995 and the squeezing parameter to 1.5, a Fisher information per photon of 31.5 $\rm{rad}^{-2}$ can be achieved in the near future (see Fig. S1 in \cite{suppl}).

Finally, exploiting its unconditional and robust quantum-enhance metrological performance, we apply our scheme in a real-time phase measurement demonstration. The to-be-measured phase is set to periodically oscillate in steps between eleven $\phi_{i}$ settings, controlled by a motor-driven positioner. At each phase, we collect 0.2 s data to obtain an estimation of  $\phi_{i}$, denoted as $\phi_{i}^{est}$, which is optimized by minimizing the squared difference between the measured probabilities and their corresponding calibration curves ($p_{ij}(\phi)$ in Fig. \ref{fig:fig3}a). The estimated phases and the preset phases are plotted together in Fig. \ref{fig:fig4}a for a comparison. The phase measurement process is repeated 200 times to calculate the phase sensitivity $\Delta \phi$, which is the standard deviation of $\phi_{i}^{est}$. The experimentally measured phase sensitivity $\Delta \phi$ are shown in Fig. \ref{fig:fig4}b, which are in a good agreement with the theoretical phase sensitivity (cyan line) calculated from the Fisher information plotted in Fig. \ref{fig:fig3}a. The squeezing parameter in this test is set at 0.43(1). The best phase sensitivity of $\Delta \phi =0.002(1)$ rad is obtained at the phase setting of 0.58, which is well beyond the SNL by 3.56 dB, and the ideal 5-NOON state interferometry. Although phase super-sensitivity is achieved only within a specific range, the entire phase range could be included by using adaptive feedback measurement \cite{wiseman1995adaptive,berni2015initio,yonezawa2012quantum}.

In conclusion, by implementing a nonlinear interferometer \cite{yurke1986su}, proposed more than 30 years ago, in a double-pass configuration of stimulated squeezed-photon emission as used in Jiuzhang \cite{zhong2021phaseprogrammable}, we have demonstrated unconditional quantum metrological advantage in phase sensing. Our method has a clear pathway to scale up, and is robust to external photon loss, thus opens a promising way to practical quantum metrology applications in the ultralow photon flux regime, such as measurement of light-sensitive materials \cite{wolfgramm2013entanglementenhanced,taylor2013biological,ono2013entanglementenhanced,israel2014supersensitive,waldchen2015lightinduced,taylor2016quantum,perarnau2021weakly,you2021scalable,casacio2021quantumenhanced}.

Note: After completing our experiment, we became aware of a related work based on direct homodyne detection of squeezed states posted on arXiv:2111.09756.

\begin{acknowledgements}
This work was supported by the National Natural Science Foundation of China, the National Key R\&D Program of China (2019YFA0308700), the Chinese Academy of Sciences, the Anhui Initiative in Quantum Information Technologies and the Science and Technology Commission of Shanghai Municipality.

\end{acknowledgements}


%
 
\end{document}